\documentclass[aps,prl,reprint,amsmath,amssymb,twocolumn]{revtex4}

\usepackage{graphicx}
\usepackage{amssymb}
\usepackage{amsmath}
\usepackage{dcolumn}
\usepackage{bm}
\usepackage{mathrsfs}
\usepackage{color}
\usepackage{hyperref}

\begin{document}

\title{Single photons from dissipation in coupled cavities}

\author{H. Flayac}
\author{V. Savona}
\affiliation{Institute of Physics, \'{E}cole Polytechnique F\'{e}d\'{e}rale de Lausanne EPFL, CH-1015 Lausanne, Switzerland}

\begin{abstract}
We propose a single photon source based on a pair of weakly nonlinear optical cavities subject to a one-directional dissipative coupling. When both cavities are driven by mutually coherent fields, sub-poissonian light is generated in the target cavity even when the nonlinear energy per photon is much smaller than the dissipation rate. The sub-poissonian character of the field holds over a delay measured by the inverse photon lifetime, as in the conventional photon blockade, thus allowing single-photon emission under pulsed excitation. We discuss a possible implementation of the dissipative coupling relevant to photonic platforms.
\end{abstract}
\maketitle

\section{Introduction}
Single photon sources \cite{Lounis2005,Eisaman2011} are fundamental building blocks for quantum information protocols. Current realizations based on blockade mechanisms \cite{Paul1982} unavoidably require a strong optical nonlinearity. They are usually engineered with such systems as quantum dots \cite{Michler2000,Santori2001,Ding2016,Somaschi2016,Schlehahn2016}, diamond color centers \cite{Kurtsiefer2000}, superconducting circuits \cite{Lang2011} or trapped atoms \cite{McKeever2004}. Although the degree of control over these systems is steadily improving, they basically operate at cryogenic temperatures and (or) imply significant fabrication challenges, particularly with respect to integration and scalability in future photonic platforms. On the other hand, nondeterministic sources relying on heralding protocols are now operating at room temperature in Silicon \cite{Davanco2012,Azzini2012,Collins2013,Li2011,Spring2013}, but they require a significant input power to trigger the four wave mixing mechanism.

The unconventional photon blockade (UPB) was proposed as a novel paradigm to produce sub-poissonian light in presence of a very weak nonlinearity \cite{Liew2010}. The seminal system consists of a pair of coherently coupled cavities embedding a Kerr nonlinear medium, where one of the cavities is driven by a classical source \cite{Liew2010,Bamba2011a,Bamba2011,Flayac2013}. It was then extended to Jaynes-Cummings \cite{Bamba2011a}, optomechanical \cite{Savona2013} or bimodal cavity \cite{Majumdar2012} systems and would be feasible in superconducting circuits \cite{Eichler2014}, or optimized silicon photonic crystal platform \cite{Ferretti2013,Flayac2015}. It was shown that UPB essentially originates from a quantum interference mechanism \cite{Liew2010,Bamba2011a}, that arises even when the nonlinear energy per photon $U\ll\kappa$, where $\kappa$ is the cavity photon loss rate. UBP was shown to be within reach of an optimized silicon photonic crystal platform \cite{Ferretti2013,Flayac2015}, where it could lead to a new class of highly integrable, ultralow-power, passive single photon sources. However, the coherent mode coupling -- with rate $J$ -- results a detrimental oscillation of the delayed two-photon correlations $g^{(2)}(\tau)$, thus restricting the sub-poissonian behavior delays shorter than $1/J\ll1/\kappa$, under the required optimal antibunching conditions \cite{Bamba2011a}. As a consequence, UPB is suppressed under pulsed excitation, as the antibunched portion of the time-dependent field contributes minimally to the emitted pulse of duration $1/\kappa$ \cite{Flayac2015}. Filtering the output pulse through a narrow time gate was shown to improve photon antibunching at the expense of a reduced photon rate \cite{Flayac2015}. Alternative schemes \cite{Kyriienko2014} rely on a strong auxiliary driving field, thus departing from the desired low-power operation.

UPB can be understood in terms of Gaussian squeezed states \cite{Lemonde2014}. For any coherent state $\left| \alpha  \right\rangle$, there exists an optimal squeezing parameter $\xi$ that minimizes the two-photon correlation $g^{(2)}(0)$, which can be made vanishing for weak driving field. In the UPB scheme, the two couple modes bring enough flexibility to tune the $\alpha$ and $\xi$ values of the target mode independently \cite{Bamba2011}. A more effective approach would then consist in pipelining two subsystems, one which provides the squeezing $\xi$ and the other that induces the corresponding optimal displacement $\alpha$.

\begin{figure}[ht]
\includegraphics[width=0.40\textwidth,clip]{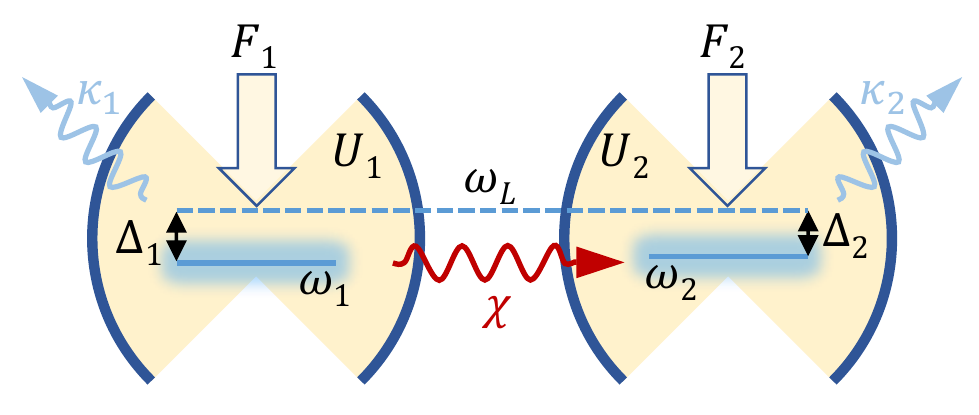}\\
\caption{(Color online) Scheme of the proposed system: Two nonlinear and dissipative optical resonators are driven by mutually coherent fields of same frequency $\omega_L$ but with different complex amplitudes $F_{1,2}$. The one-directional dissipative coupling from cavity 1 to cavity 2 occurs at a rate $\chi$.}
\label{Fig1}
\end{figure}

In this paper, we develop such an approach by investigating a scheme where two optical resonators are linked via a dissipative -- i.e. one-directional -- coupling \cite{Carmichael1993,Gardiner1993,Gardiner1994}. The nature of such coupling allows independent tuning of the field squeezing and displacement in the second cavity. The most significant advance brought by the one-directional coupling however, is the absence of a normal-mode energy splitting. This removes the oscillations typical of UPB, so that the condition $g^{(2)}(\tau)<1$ is fulfilled for delays longer than the cavity lifetime, and pulsed operation becomes naturally possible as in the conventional photon blockade.

\section{The Model}
We consider two driven Kerr resonators with dissipative (one-directional) coupling from the first to the second cavity as sketched in Fig.\ref{Fig1}. Our goal is to find a regime of parameters for which cavity 2 -- from now on denoted target cavity -- displays sub-poissonian photon statistics. In the frame rotating at the frequency $\omega_L$ of the driving fields, the system Hamiltonian reads
\begin{equation}\label{H}
  \hat {\cal{H}} = \sum\limits_{j = 1,2} {\left[ -{{\Delta_j}\hat a_j^\dag {{\hat a}_j} + U_j\hat a_j^\dag \hat a_j^\dag {{\hat a}_j}{{\hat a}_j}} + F_j^*\hat a_j + F_j\hat a_j^\dag \right]}\,,
\end{equation}
where $F_j$ ($j=1,2$) are the complex driving field amplitudes for each cavity, $\Delta_{j}=\omega_{j}-\omega_L$ are the cavity mode detunings, and $U_{j}$ are the strengths of the Kerr nonlinearities. The open system dynamics obeys the quantum master equation
\begin{equation}\label{rhot}
  i\frac{{\partial \hat \rho }}{{\partial t}} =  \left[ {\hat {\cal{H}},\hat \rho } \right] - {\frac{{{i}}}{2}\sum\limits_{j = 1,2} \kappa _j\hat {\cal{D}}\left[ {{{\hat a}_j}} \right]\hat \rho}  + i\chi \hat {\cal{D}}\left[ {{{\hat a}_1},{{\hat a}_2}} \right]\hat \rho\,,
\end{equation}
where $\hat{\cal{D}}\left[ {{{\hat a}_j}} \right]\hat \rho = \{\hat a_j^\dag {{\hat a}_j},\hat \rho\} - 2{{\hat a}_j}\hat \rho \hat a_j^\dag$ describe the dissipation into the environment at rates $\kappa_j$, and $\hat{\cal{D}}\left[ {{{\hat a}_1},{{\hat a}_2}} \right]\hat \rho = [ {{{\hat a}_1}\hat \rho ,\hat a_2^\dag } ] + [ {{{\hat a}_2},\hat \rho \hat a_1^\dag } ]$ models the dissipative coupling at a rate $\chi  = \sqrt {\eta {\kappa _1}{\kappa _2}}$. Here, we have defined a transfer efficiency $\eta \in [0,1]$ \cite{Gardiner2004}, relating the coupling to the dissipation rates.

Before directly solving Eq. (\ref{rhot}), it is useful to study the system in the limit of weak driving fields $F_{1,2} \rightarrow 0$. In this limit, analytical expressions for the various expectation values can be obtained by assuming pure states and restricting to the $n\le2$ photon manifold, as was also done in Refs. \onlinecite{Bamba2011a,Savona2013,Flayac2013}. Details of this analysis are reported in the Appendix A. In that framework, the steady state two-photon correlations of the target cavity approximates to
\begin{eqnarray}
g_{2}^{(2)}(0)=\frac{\langle\hat{a}^{\dag}_{2}\hat{a}^{\dag}_{2}\hat{a}_{2}\hat{
a}_{2}\rangle}{\langle\hat{a}^{\dag}_{2}\hat{a}_{2}\rangle^2}\simeq2\frac{|c_{02}|^2}{|c_{01}|^4}\label{g2}\,,
\end{eqnarray}
where $|c_{01}|^2$ and $|c_{02}|^2$ are the probabilities of having zero photons in cavity 1 and, respectively, 1 and 2 photons in the target cavity [see Eqs.(\ref{c01},\ref{c02})]. By requiring $c_{02}=0$, one obtains a condition for an optimal sub-poissonian behavior, $g_2^{(2)}(0) \simeq 0$. This optimal condition can be met, provided that the driving fields fulfill
\begin{equation}\label{F1opt}
  {F_1}\left| {_{\rm opt}} \right. = i{F_2}\frac{{\tilde \Delta {{\tilde U}_1} \pm \sqrt { - {U_1}{{\tilde U}_1}\tilde \Delta {{\tilde \Delta }_2}} }}{{\left( {\tilde \Delta  + {U_1}} \right)\chi }}\,,
\end{equation}
where we defined $\tilde \Delta_{1,2}=-\Delta_{1,2}-i\kappa_{1,2}/2$, $\tilde \Delta  = {{\tilde \Delta }_1} + {{\tilde \Delta }_2}$, ${{\tilde U}_1} = {{\tilde \Delta }_1} + {U_1}$, and assuming $F_2\in {\mathbb{R}^{+}}$ without loss of generality. Eq.\eqref{F1opt} reveals several interesting features: (i) ${F_1}\left| {_{\rm opt}} \right.$ needs to carry the proper magnitude and phase. (ii) ${F_1}\left| {_{\rm opt}} \right.$ depends linearly on $F_2$, which cannot therefore be set to zero. Indeed, an undriven target cavity would simply act as a spectral filter \cite{Flayac2014}, thus essentially recovering the single mode statistics. (iii) The optimal field amplitude doesn't depend on $U_2$, which can therefore be set arbitrarily small. If under the assumption that the sub-poissonian character originates from the optimal squeezing mechanism described in Ref.\cite{Lemonde2014}, then this feature hints at the fact that cavity 1 is here the main source of squeezing. (iv) In the case where $U_1=0$, an optimal value of the driving field is still well defined but results in a vanishing occupation of the target cavity. (v) We made no assumptions on the value of $U_j$, which may be set arbitrarily smaller or larger than the loss rates $\kappa_j$.

\section{Results and discussion}
\begin{figure}[ht]
\includegraphics[width=0.5\textwidth,clip]{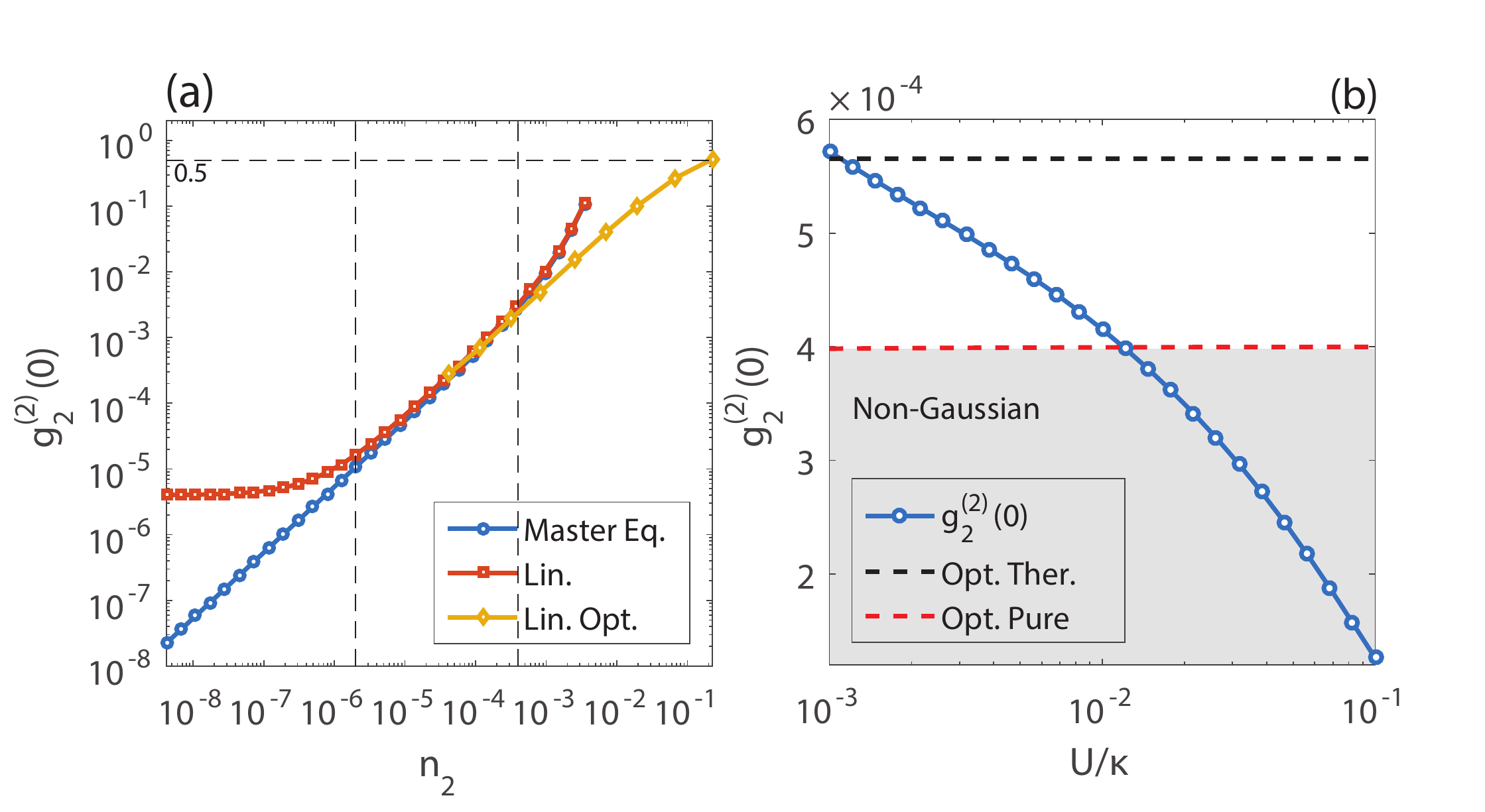}\\
\caption{(Color online) (a) Target cavity two-photon correlations $g_2^{(2)}(0)$ versus its occupation $n_2$. Here $U=10^{-3}\kappa$, $F_1=F_1|_{\rm{opt}}$ and $\chi=\kappa$. Blue line: exact master equation. Red line: linearized model. Yellow line: linearized model where $F_1$ was obtained for each value of $n_2$ from numerical minimization of $g_2^{(2)}(0)$. (b) $g_2^{(2)}(0)$ versus $U$ at fixed occupation $n_2=10^{-4}$. The dashed black and red line denote the thermal and pure state Gaussian boundaries set in Ref. \onlinecite{Lemonde2014}.}
\label{Fig2}
\end{figure}

From now on, we will consider the case of cavities with equal loss rates $\kappa_1=\kappa_2=\kappa$ and and nonlinearities $U_1=U_2=U$. We solve numerically Eq. \eqref{rhot} in the stationary limit, on a Hilbert space truncated to include $N_{\rm{max}}$ quanta per mode. As a figure of merit, in Fig.\ref{Fig2}(a), we show the two-photon correlation $g_2^{(2)}(0)$ for the target cavity, as a function of its average occupation $n_2$ (blue line). We assumed for this calculation the most interesting regime of weak nonlinearity compatible with Silicon photonic crystal cavities where $U=10^{-3}\kappa$ \cite{Flayac2015}. We additionally assumed $\Delta_1=\Delta_2=0$ for simplicity. Since $U\ll\kappa$, the optimal condition \eqref{F1opt} approximately reduces to
\begin{equation}\label{F1opts}
  {F_1}\left| {_{\rm opt}} \right. \simeq i{F_2}\frac{\kappa }{{2\chi }}
\end{equation}

For increasing driving field amplitudes, the value of $N_{\rm{max}}$ required for convergence becomes exceedingly large. To extend the range of accessible $n_2$ values, we linearize with respect to the mean-field solution (see Appendix B). The result is plotted in Fig.\ref{Fig2}(a) (red line) and matches perfectly with the full quantum treatment from $n_2>10^{-6}$ where the mean field dominates over fluctuations. The optimal two-photon correlation behaves linearly as a function of $n_2$, with the exception of larger occupancies where a nonlinear increase in $g_2^{(2)}(0)$ is displayed. This behavior resides in the limited range of validity of Eq.\eqref{F1opt} which loses accuracy as the 3 photon probability rises. For the largest values of $n_2$ we therefore searched for the optimal parameters numerically, using the amplitude and phase of $F_1$ as free parameters. The value $g_2^{(2)}(0)\leqslant0.5$ -- considered as an upper bound for single-photon emission -- is reached for a remarkably high occupancy $n_2\simeq0.25$ (yellow line) for such a weakly nonlinear system. We note that in the presence of a thermal background, e.g. if microwave photons \cite{Eichler2014} are envisaged, the system would display a value of $g_2^{(2)}(0)=2$ in the limit of vanishing driving fields. The function would therefore present a minimum at finite driving amplitude.

By assuming a linewidth $\kappa=1$ $\mu$eV of state-of-the-art photonic crystal cavities, we can extract a maximum emission rate as high as ${\cal{R}}=n_2\kappa/\hbar=380$ MHz. The corresponding intracavity power at zero detuning for cavity resonances $\hbar\omega_{1,2}=0.8$ eV is $P_{\rm in}=\omega_c (F_{1}+F_{2})/\hbar=15.5$ pW given $F_2\simeq\sqrt{n_2}\kappa = 2 F_1$. The real input power can be estimated to $50 \times P_{\rm in}=778$ pW when taking into account a conservative value for the in-coupling efficiency \cite{Dharanipathy2014}. This value is about 30 times smaller than the typical input power required for single photon operation with quantum dots \cite{Michler2000}. 

It was shown \cite{Lemonde2014} that, under the assumption that the state is Gaussian, a lower bound on $g^{(2)}(0)$ exists. In particular, for mean occupancies $n \ll 1$, this bound is given by $g^{(2)}(0)|_{\rm{p}}\simeq4|\langle\hat a\rangle|^2$ for a pure displaced-squeezed state, and by $g^{(2)}(0)|_{\rm{th}}\simeq8\sqrt{\bar{n}_{\rm{eff}}}$ for a corresponding thermal (i.e. mixed) state with mean occupation $\bar{n}_{\rm{eff}}$. For a general mixed state, we can define an effective thermal occupation $\bar{n}_{\rm{eff}}=(1/{\rm Tr} {{{\hat \rho }^2}} - 1)/2\ll1$, which then roughly measures the degree of mixedness. We show in Fig.\ref{Fig2}(b) the computed $g_2^{(2)}(0)$ as a function of $U$ (blue line) at constant occupation $n_2=10^{-4}$ (where Eq.(\ref{F1opt}) holds), and compare it to the pure and thermal limits (dashed lines) that are independent of $U$ for given values of $n_2$ and $\kappa$. Photons in the target cavity achieve a value of $g^{(2)}(0)$ lying below the thermal limit and, from $U/\kappa>10^{-2}$, crossing the pure state boundary. In this case, the state departs from a Gaussian state, which was checked by identifying negative Wigner distribution areas (not shown).

\begin{figure}[ht]
\includegraphics[width=0.5\textwidth,clip]{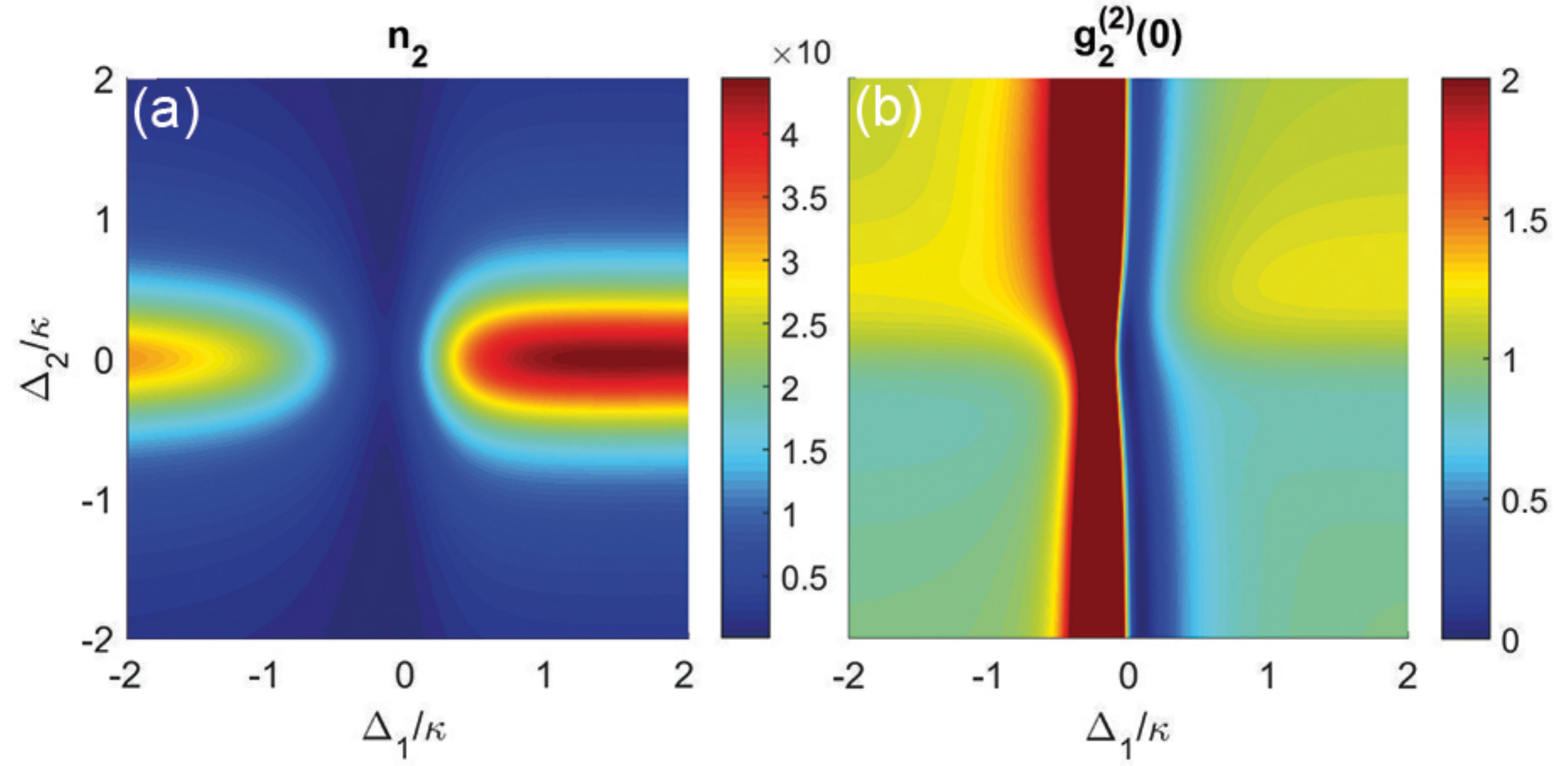}\\
\caption{(Color online) Color maps of (a) the target cavity occupation $n_{2}$ and (b) two-photon correlation $g_{2}^{(2)}(0)$, computed as a function of the cavity detunings. Here $U=10^{-3}\kappa$, $\chi=\kappa$ and we set the optimal condition (\ref{F1opt}) at $\Delta_{1,2}=0$ and $n_2=10^{-4}$.}
\label{Fig3}
\end{figure}

We have studied the impact of variable detunings $\Delta_{1,2}$ when the other parameters are fixed. The results for $n_2$ and $g_2^{(2)}(0)$ are presented in Fig.\ref{Fig3}. The panel (a) shows that the occupation vanishes for small detunings. This is due to destructive interference between the input from cavity 1 and the field driving the target cavity. In particular, under the condition $F_2=-i \chi F_1/\tilde \Delta_1$, the coefficient $c_{01}$ is suppressed therefore favoring photon pairs (see Appendix A). As shown in Fig.\ref{Fig3}(b), in the region $\Delta_1<|\kappa/2|$, we observe both a strong bunching up to  $g_2^{(2)}(0)=30$ (red areas) or strong antibunching (blue areas) where the optimal antibunching condition holds. As already discussed, antibunching results from the interplay between the squeezing brought by cavity 1 and the field displacement induced by the driving field on the target cavity \cite{Lemonde2014}. The results are essentially unchanged when $U_2=0$ as dictated by Eq.\eqref{F1opt}.

\begin{figure}[ht]
\includegraphics[width=0.45\textwidth,clip]{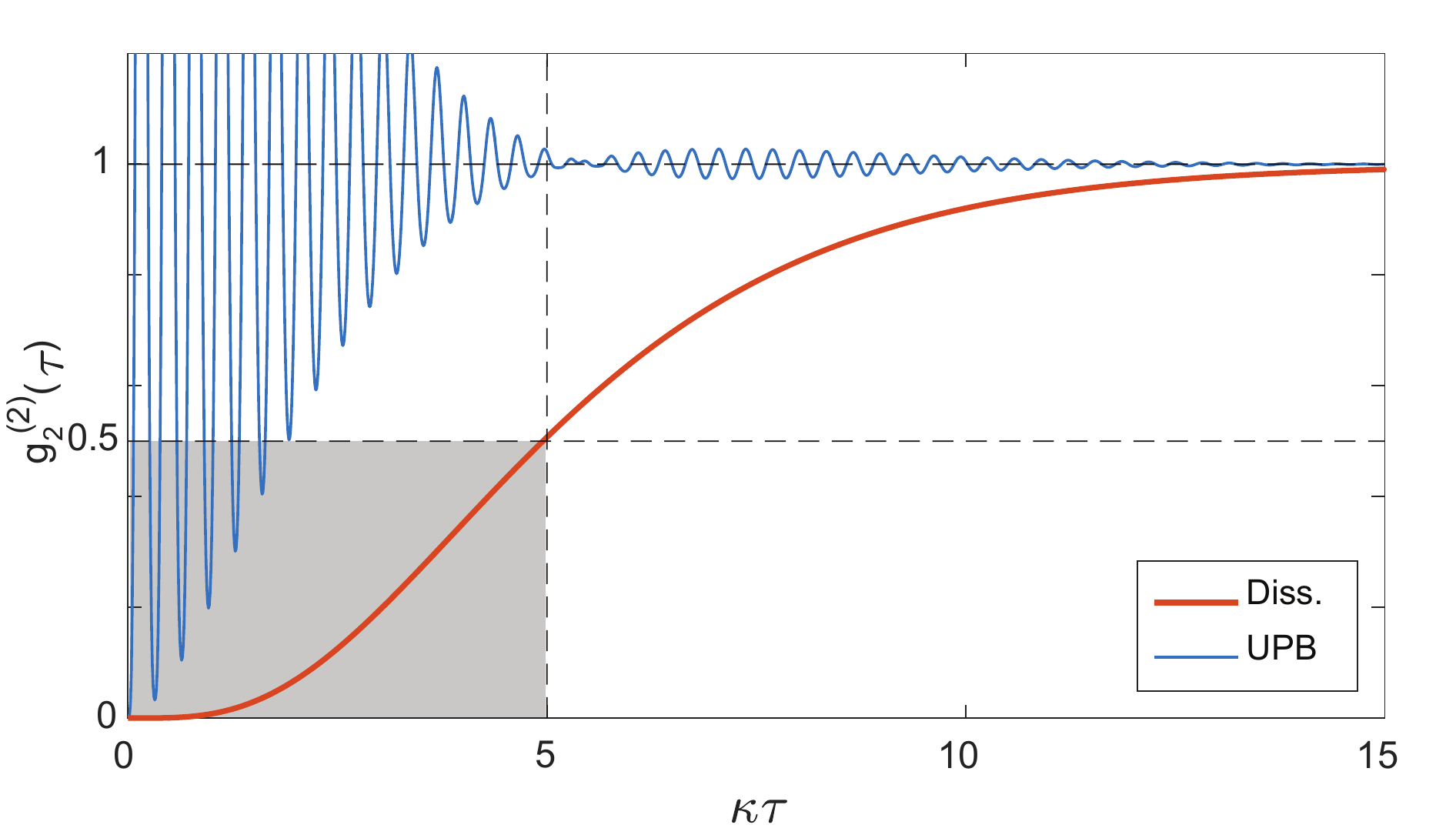}\\
\caption{(Color online) Delayed two-photon correlation function $g_2^{(2)}(\tau)$ (red line) computed in the steady state regime under continuous wave driving, for the target cavity at $n_2=10^{-4}$, $U=10^{-3}\kappa$ and $\chi=\kappa$. The gray area highlights the single photon regime. The blue line shows the oscillating UPB counterpart obtained for the same value of $U$ requiring $J=19.6\kappa$, $\Delta_j=0.29\kappa$ and $F_1=0$.}
\label{Fig4}
\end{figure}

As discussed above, the UPB scheme displays antibunching only for values of the time delay smaller than $1/J \ll 1/\kappa$ \cite{Liew2010,Bamba2011a,Flayac2015}, thus preventing simple operation under pulsed input. This is ultimately due to the normal-mode energy splitting in the spectrum of the two-resonators, of the order of $2J$. The dissipative coupling overcomes this difficulty, as the normal-mode splitting is absent and the emitted photons are characterized by the spectrum of the target cavity (see Fig.A\ref{FigS4}). We show in Fig.\ref{Fig4} the $g_2^{\left( 2 \right)}(\tau)$ function computed at steady state for the optimal parameter values (red line), and compare it to the UPB result (blue line) for the same value of $U$. In the dissipative case, the antibunching survives over $\tau>1/\kappa$ and oscillations are absent. The single photon regime, defined by $g^{(2)}_2(\tau)<0.5$, is preserved over the shaded time frame and behaves similarly to conventional sources \cite{Paul1982}.

We studied the pulsed regime in more detail by a direct time integration of Eq.\eqref{rhot} where we assumed input Gaussian pulses $F_j\exp[-(t-t_{j0})^2/\sigma_t^2]$. A key quantity in assessing the single-photon emission under pulsed excitation is the two-photon correlation averaged over two times \cite{Flayac2015}
\begin{equation}\label{g2pulse}
  g_{\rm{pulse}}^{(2)} = \frac{{\int {G_2^{(2)}\left( {{t_1},{t_2}} \right)d{t_1}d{t_2}} }}{{\int {{n_2}\left( {{t_1}} \right){n_2}\left( {{t_2}} \right)d{t_1}d{t_2}} }}\,,
\end{equation}
where $G_2^{(2)}({t_1},{t_2})=\langle {\hat a_2^\dag(t_1) \hat a_2^\dag(t_2) {{\hat a}_2}(t_2){{\hat a}_2}(t_1)} \rangle$ and $n_2(t)= \langle\hat a_2^\dag(t) {{\hat a}_2}(t)\rangle$. For optimal single-photon operation, the duration of the excitation pulses should be optimized so to be shorter than the sub-poissonian time window (see Fig.\ref{Fig4}), while allowing enough time for the buildup of the squeezing (see Fig.A\ref{FigS0}). A suitable delay $\Delta t=t_{02}-t_{01}=1.5/\kappa$ between the two pulses (see Appendix C) has also been introduced here to circumvent the onset of strong bunching in the earliest part of the output pulse.

\begin{figure}[ht]
\includegraphics[width=0.5\textwidth,clip]{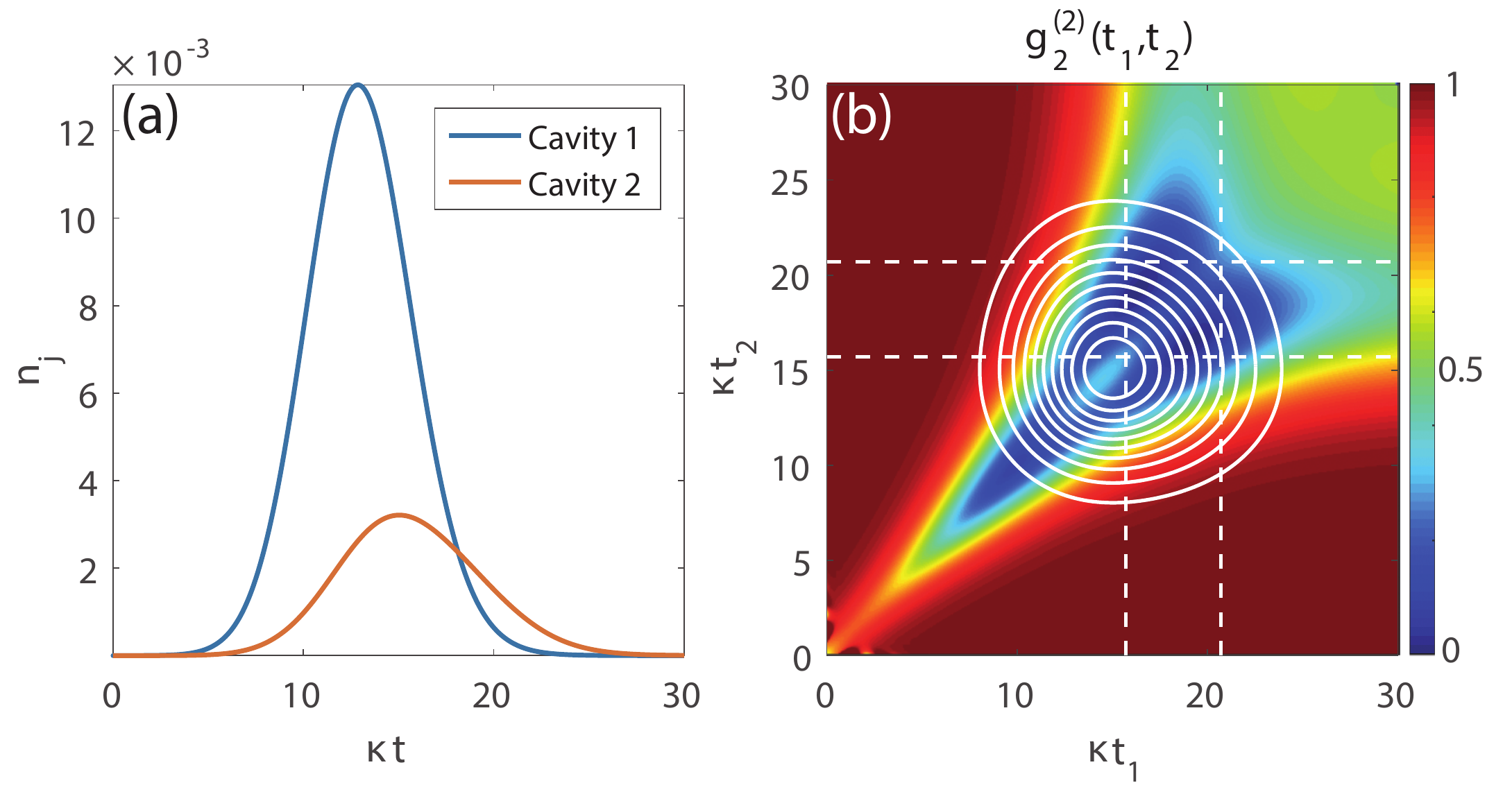}\\
\caption{(Color online) Pulsed regime: (a) Time dependent cavity occupation. (b) Two-time two-photon correlation function $g^{(2)}_2(t_1,t_2)$. The quantity $n_2(t_1,t_2)$ is displayed as a contour plot. The dashed-white lines denote a time-gate window resulting in $g_{\rm{pulse}}^{(2)}<0.1$. The parameters are $U=10^{-1}\kappa$, $\chi=\kappa$, $\sigma_t=5/\kappa$, $\Delta_{1,2}=0$, $F_2=0.1\kappa$, $F_1=F_1|_{\rm{opt}}$ and $\Delta t=1.5/\kappa$.}
\label{Fig5}
\end{figure}

We show in Fig.\ref{Fig5}(a) the computed cavity occupations $n_{1,2}(t)$. Here, the target cavity has an average occupation of $n_{\rm{pulse}}=\int n_2(t)dt\simeq3\times10^{-2}$. Fig.\ref{Fig5}(b) shows the corresponding two-time correlation function $g^{(2)}_2(t_1,t_2)$. The contour plot highlights the occupation of the target cavity $n_2(t_1,t_2)=\sqrt{n_2(t_1) n_2(t_2)}$, which peaks well inside the sub-poissonian portion of the plot. For the present case, we obtained $g_{\rm{pulse}}^{(2)}\simeq0.3$. Single photon operation may be enhanced via an optimal pulse shaping (a task beyond the scope of this study). Further enhancement may be obtained through time-gating the output pulse, as already suggested for the UPB \cite{Flayac2015}. If one applies a time gate of duration $\Delta T=5/\kappa$, highlighted by the dashed lines in Fig.\ref{Fig5}(b), the two-photon correlation is reduced to $g_{\rm{pulse}}^{(2)}<0.1$ while preserving an average occupation of $n_{\rm{pulse}}\simeq 10^{-2}$. In line with the steady-state discussion and for the parameters we chose here, which fit in the requirements of condition \eqref{F1opt}, we can estimate a single photon rate of ${\cal R}=2.3$ MHz if we assume pulses delayed by $20/\kappa=13$ ns. Note that this rate could easily be increased by one order of magnitude by considering a numerically optimized pump amplitudes as in Fig.\ref{Fig2}(a) (yellow curve).

The dissipative coupling considered so far can be implemented through an intermediate coupling element, which may be a waveguide or a third optical resonator, as investigated in Ref.\cite{Metelmann2015}. The coupling element acts as an engineered reservoir, effectively generating the quantum interference required for the one-directional transmission. In the case of a third cavity, the corresponding Hamiltonian reads
\begin{eqnarray}\label{H3}
\nonumber  \hat {\cal{H}} &=&  \sum\limits_{j = 1}^3 {\left[ { - {\Delta _j}\hat a_j^\dag {{\hat a}_j} + {U_j}\hat a_j^\dag \hat a_j^\dag {{\hat a}_j}{{\hat a}_j} + F_j^*{{\hat a}_j} + {F_j}\hat a_j^\dag } \right]}\\
   &+& \sum\limits_{j \ne k = 1}^3 {\left[ {{J_{jk}}\hat a_j^\dag {{\hat a}_k} + J_{jk}^*\hat a_k^\dag {{\hat a}_j}} \right]}
\end{eqnarray}
where the auxiliary mode is not driven, i.e. $F_3=0$, and is ideally characterized by a large dissipation rate $\kappa_3\gg\kappa_{1,2}$. $J_{jk}$ describe coherent photon hopping amplitudes. This system is well approximated by Eq. \eqref{rhot} under the conditions $J_{12} = i{\chi }/{2}$ and $J_{23} = J_{31}=\sqrt{-iJ_{12}\kappa_3/2}$ \cite{Metelmann2015}. It requires a complex-valued $J_{12}$, currently feasible e.g. using waveguide delay lines \cite{Hafezi2011}. The directionality of the coupling can be tested numerically (see Appendix D). In presence of the auxiliary resonator, the optimal antibunching condition is displaced in the parameter space. We have identified a new condition by running a steady state optimization with $|F_1|$ and $\phi_1$ as free parameters, and setting $F_2=0.1\kappa$, $U_j=10^{-3}\kappa$, $\Delta_j=0$, and $\kappa_3=10\kappa$. We obtained a value of $g^{(2)}_2(0)=3.8\times10^{-2}$ for $|F_1|=6.58\times10^{-2}\kappa$ and $\phi_1\simeq0$, proving the single-photon operation.

An efficient single-photon source should be benchmarked against the current state-of-the-art, represented by quantum emitters in resonant cavities \cite{Michler2000,Santori2001,Kurtsiefer2000,McKeever2004,Lang2011,Ding2016,Somaschi2016,Schlehahn2016}, and heralded sources\cite{Li2011,Davanco2012,Azzini2012,Collins2013,Spring2013}. Recent advances have led to close-to-ideal single-photon operation for both schemes. However, quantum emitters and heralded source respectively require cryogenic temperature and high input power. The present proposal brings a significant advantage in that it naturally operates at ultra-low power, in the nW range as estimated above for a photonic crystal cavity. This must be compared to the 24 nW of Ref.\cite{Michler2000} and to the mW range of heralded sources \cite{Azzini2012}. Expected single-photon rates are in the MHz range for the three schemes \cite{Azzini2012,Ding2016,Flayac2015}. Photon purity (i.e. the value of $g^{(2)}(0)$) can be made here arbitrarily large, as seen in Fig.\ref{Fig2}(a).
Moreover here, photons are emitted within the narrow spectrum of the target cavity (see Fig.A\ref{FigS4}(a)). Within the assumptions of our model, the indistinguishability degree amounts to $99.95\%$ (see Appendix E). Such a high value would obviously be reduced in the presence to pure dephasing or fluctuations in the driving fields. This incidentally represents a second advantage -- in addition to pulsed operation -- of the present scheme on the original UPB, where instead photons are emitted over the spectrum of the normal modes of the two cavities. The probabilistic emission character remains a limitation of both UPB and cascaded schemes. We are confident however that this difficulty may be overcome soon by devising new schemes based on the present dissipative coupling paradigm.

\section{Conclusion}
We have proposed a scheme for a single-photon source operating under weak nonlinearity and relying on dissipative, one-directional coupling between two optical cavities. Such approach enables single-photon generation over pulsed excitation, thus overcoming the main limitation of the unconventional photon blockade. We have proposed a three-cavity configuration that enables the one-directional coupling and may be realized on several platforms, including weakly nonlinear photonic crystal cavities, coupled ring resonators. The scheme could be generalized to several cascaded optical cavities, aiming at suppressing the $n$-photon probabilities to enhance the single-photon operation or pair production.

\begin{acknowledgments}
The authors acknowledge fruitful discussions with D. Gerace and M. Minkov.
\end{acknowledgments}

\
\appendix*
\
\setcounter{figure}{0}
\section{Appendix A: Weak pump limit}\label{AppA}
\begin{figure}[ht]
\renewcommand{\figurename}{Fig.A}
\includegraphics[width=0.5\textwidth,clip]{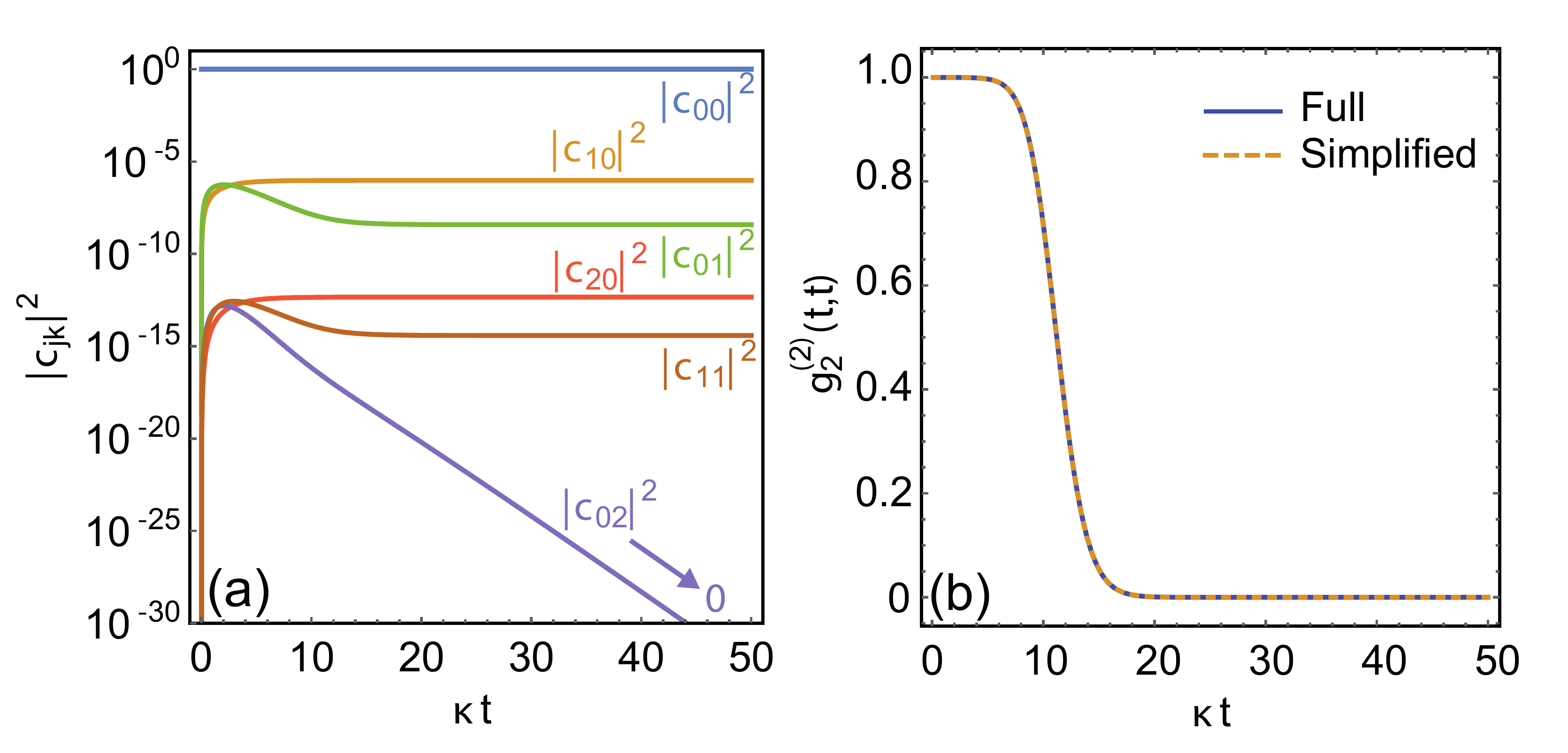}\\
\caption{(Color online) (a) Log scale wavefunction coefficients evolution in the 2 photons subspace. (b) Target cavity two photon correlation evolution using the full $n_2$ expression (blue line) or neglecting the $c_{02}$, $c_{11}$ contribution and the Eqs.(\ref{c10},\ref{c01}) underlined terms (dashed red line). Here $F_2=10^{-3}\kappa$ and $F_1=F_1|_{{\rm opt}}$. The other parameters are those of Fig.2(a) of the main text.}
\label{FigS0}
\end{figure}

Given the system Hamiltonian
\begin{equation}\label{H}
  \hat {\cal{H}} = \sum\limits_{j = 1,2} {\left[ -{{\Delta_j}\hat a_j^\dag {{\hat a}_j} + U_j\hat a_j^\dag \hat a_j^\dag {{\hat a}_j}{{\hat a}_j}} + F_j^*\hat a_j + F_j\hat a_j^\dag \right]},
\end{equation}
we can express the time-dependent quantum state as an expansion on the basis of occupation number eigenstates.
In the limit of weak driving fields $F_{1,2}\rightarrow0$, it is then possible to retain only terms in this expansion, whose coefficients depend to leading order in the driving field amplitudes as ${\cal{O}}( {F_1^j F_2^k})$ with $j+k\le2$. From the Schr\"odinger equation, it can be easily inferred that the coefficient $c_{jk}$ depends exactly as ${\cal{O}}( {F_1^j F_2^k})$ to leading order. Hence, this weak driving limit coincides with approximating the time-dependent state as
\begin{eqnarray}\label{psi}
\left| \psi \right\rangle  &\simeq& {c_{00}}\left| {00} \right\rangle + {c_{10}}\left| {10} \right\rangle + {c_{01}}\left| {01} \right\rangle\\
\nonumber                  &+& {c_{11}}\left| {11} \right\rangle + {c_{20}}\left| {20} \right\rangle + {c_{02}}\left| {02} \right\rangle
\end{eqnarray}
where, $\left| jk \right\rangle$ denotes a Fock state with $j$ photons in the first cavity and $k$ photons in the second one. The equations governing the time-dependence of the coefficients are found from the solution of the Schr\"{o}dinger equation $\tilde {\cal H}\left|\psi\right\rangle=i\hbar{\partial _t}\left| \psi  \right\rangle$ written for the non-hermitian Hamiltonian
\begin{equation}\label{HNH}
\tilde {\cal{H}} = \hat{\cal{H}} - \frac{i}{2}\sum\limits_{j = 1,2} {{\kappa _j}\hat a_j^\dag {{\hat a}_j}} - i\chi {{\hat a}_1}{\hat a_2^\dag}
\end{equation}
where the first term stands for the cavity losses and the second one for the one-directional transmission. We obtain the following coupled set of equations for the coefficients $c_{jk}(t)$
\begin{eqnarray}
  i\dot c_{00} &=& {F_1^*}{c_{10}}+ {F_2^*}{c_{01}}\\
  \label{dc10}
  i\dot c_{10} &=& {F_1}{c_{00}} + {{\tilde \Delta }_1}{c_{10}} + \underline {{F_1^*}\sqrt 2 {c_{20}}}  + \underline {{F_2^*}{c_{11}}}\\
  \label{dc01}
\nonumber i\dot c_{01} &=& {F_2}{c_{00}} + {{\tilde \Delta }_2}{c_{01}} + \underline {{F_2^*}\sqrt 2 {c_{02}}}  + \underline {{F_1^*}{c_{11}}} - i\chi {c_{10}}\\ \\
  i\dot c_{20} &=& {F_1}\sqrt 2 {c_{10}} + 2\left( {{U_1} + {{\tilde \Delta }_1}} \right){c_{20}}\\
  \label{c01c02c11}
  i\dot c_{02} &=& {F_2}\sqrt 2 {c_{01}} + 2\left( {{U_2} + {{\tilde \Delta }_2}} \right){c_{02}} - i\chi \sqrt 2 {c_{11}}\\
\nonumber  i\dot c_{11} &=& {F_1}{c_{01}} + {F_2}{c_{10}} + \left( {{{\tilde \Delta }_1} + {{\tilde \Delta }_2}} \right){c_{11}} - i\chi \sqrt 2 {c_{20}}\\
\end{eqnarray}
Note that the underlined terms in Eqs.(\ref{dc10},\ref{dc01}) are of third order in $F_{1,2}$, according to the criterion derived above, and are therefore negligible within the present approximation. In the steady state where $\dot{c}_{jk}(t)=0$, imposing the normalization condition $c_{00}=1$ and solving the resulting equations for the $c_{jk}$ coefficients iteratively, we obtain the explicit expressions
\begin{eqnarray}
  \label{c10}
  {c_{10}} &=&  - \frac{{{F_1}}}{{{{\tilde \Delta }_1}}}\\
  \label{c01}
  {c_{01}} &=&  - \frac{{{F_2}}}{{{{\tilde \Delta }_2}}} - i\chi \frac{{{F_1}}}{{{{\tilde \Delta }_1}{{\tilde \Delta }_2}}}\\
  \label{c11}
  {c_{11}} &=& \frac{{{F_1}{F_2}}}{{{{\tilde \Delta }_1}{{\tilde \Delta }_2}}} + \frac{{iF_1^2\left( {{{\tilde \Delta }_1} + {{\tilde \Delta }_2} + {U_1}} \right)\chi }}{{{{\tilde \Delta }_1}{{\tilde \Delta }_2}\left( {{{\tilde \Delta }_1} + {{\tilde \Delta }_2}} \right)\left( {{{\tilde \Delta }_1} + {U_1}} \right)}}\\
  \label{c20}
  {c_{20}} &=& \frac{{F_1^2}}{{{{\tilde \Delta }_1}\left( {{{\tilde \Delta }_1}} + {U_1} \right)\sqrt 2 }}\\
  \label{c02}
  {c_{02}} &=&
  \frac{{F_2^2}}{{{{\tilde \Delta }_2}\left( {{{\tilde \Delta }_2} + {U_2}} \right)\sqrt 2 }} \\
\nonumber   &+& \frac{{2i{F_1}{F_2}\sqrt 2 \left( {{{\tilde \Delta }_1} + {{\tilde \Delta }_2}} \right)\left( {{{\tilde \Delta }_1} + {U_1}} \right)\chi }}{{{{\tilde \Delta }_1}{{\tilde \Delta }_2}\left( {{{\tilde \Delta }_1} + {{\tilde \Delta }_2}} \right)\left( {{{\tilde \Delta }_1} + {U_1}} \right)\left( {{{\tilde \Delta }_2} + {U_2}} \right)}} \\
   \nonumber   &-& \frac{{F_1^2\left( {{{\tilde \Delta }_1} + {{\tilde \Delta }_2} + {U_1}} \right){\chi ^2}}}{{{{\tilde \Delta }_1}{{\tilde \Delta }_2}\left( {{{\tilde \Delta }_1} + {{\tilde \Delta }_2}} \right)\left( {{{\tilde \Delta }_1} + {U_1}} \right)\left( {{{\tilde \Delta }_2} + {U_2}} \right)}}
\end{eqnarray}
with the definition $\tilde \Delta_j=-\Delta_j-i\kappa_j/2$. The cavities occupations and their second order coherence functions are then computed as
\begin{eqnarray}
\label{n1}
\nonumber {n_1} &=& \langle {\hat a_1^\dag {{\hat a}_1}} \rangle = \left| c_{10} \right|^2 + \left|c_{11}\right|^2 + 2\left|c_{20}\right|^2 \simeq \left| c_{10} \right|^2 \hfill \\ \\
\label{n2}
\nonumber {n_2} &=& \langle {\hat a_2^\dag {{\hat a}_2}} \rangle = \left| c_{01} \right|^2 + \left|c_{11}\right|^2 + 2\left|c_{02}\right|^2 \simeq \left| c_{01} \right|^2 \hfill \\ \\
\label{g21}
g_1^{(2)}(0) &=& \frac{{\langle \hat a_1^\dag \hat a_1^\dag {{\hat a}_1}{{\hat a}_1}\rangle }}{{n_1^2}} \simeq 2\frac{{{{\left| {{c_{20}}} \right|}^2}}}{{{{\left| {{c_{10}}} \right|}^4}}} \\
\label{g22}
g_2^{(2)}(0) &=& \frac{{\langle \hat a_2^\dag \hat a_2^\dag {{\hat a}_2}{{\hat a}_2}\rangle }}{{n_2^2}} \simeq 2\frac{{{{\left| {{c_{02}}} \right|}^2}}}{{{{\left| {{c_{01}}} \right|}^4}}}
\end{eqnarray}
In Eqs.(\ref{n1},\ref{n2}), we resort to the fact that $c_{00}\gg c_{10}, c_{01}\gg c_{20}, c_{02}, c_{11}$ (see Fig.A\ref{FigS0}(a)). Interestingly the $c_{01}$ coefficient vanishes under the condition
\begin{equation}\label{n20}
  {F_2} =  - i\frac{\chi }{{{{\tilde \Delta }_1}}}F_1
\end{equation}
It corresponds to the condition where the laser driving of cavity 2 interferes destructively with the input $i \chi c_{10}$ from cavity 1. In that case the $\left| {02} \right\rangle$ state and therefore photon pairs are favored in the target cavity.

For single photon operation, we can now look more specifically at the conditions required for $g_2^{(2)}(0)$ to vanish namely when $c_{02}=0$. In such a case we obtain the optimal value for the cavity 1 pump amplitude
\begin{equation}\label{F1optA}
  {{F_1}\left| {_{\rm opt}} \right. = i\frac{{{F_2}\tilde \Delta {{\tilde U}_1} \pm \sqrt { - {F_2^2}{U_1}{{\tilde U}_1}\tilde \Delta {{\tilde \Delta }_2}} }}{{\left( {\tilde \Delta  + {U_1}} \right)\chi }}}
\end{equation}
where we defined $\tilde \Delta  = {{\tilde \Delta }_1} + {{\tilde \Delta }_2}$, ${{\tilde U}_1} = {{\tilde \Delta }_1} + {U_1}$. Note that we kept the pump amplitude complex all along the derivation. The optimal antibunching condition therefore requires a proper phase relation between the driving fields. Considering as above $\Delta_{1,2}=0$, $U_{1,2}=U$, $\kappa_{1,2}=\kappa$ and $F_2\in {\mathbb{R}^{+}}$ the target cavity occupation in the optimal condition reduces to
\begin{equation}\label{n2opt}
  n_2 \simeq \frac{{4F_2^2U}}{{{\kappa ^2}\left( {\kappa  - U} \right)}}
\end{equation}
and therefore strongly increases with $U$ especially when it approaches $\kappa$.

The validity of the above assumptions can be checked from Fig.A\ref{FigS0}(a) showing the analytical solutions to Eqs.(\ref{c10}-\ref{c02}) and specifically the $|c_{jk}(t)|^2$ coefficient evolution under condition (\ref{F1optA}) within the 2 photons photon subspace. It which reveals the dynamical suppression of the $c_{02}$ coefficient and the clear hierarchy between the coefficients of different manifolds separated by at least 5 orders of magnitude in such weak pump limit. In Fig.A\ref{FigS0}(b) we show the corresponding $g_2^{(2)}(t,t)$ vanishing using the full $n_2$ expression and accounting for the Eqs.(\ref{c10},\ref{c01}) underlined terms (blue line), or neglecting the $c_{02}$ and $c_{11}$ contribution and the underlined terms (dashed red line). One can see that we obtain a perfect match between the curves which insures the legitimacy of Eq.(\ref{g22}).

\section{Appendix B: Strong pump limit}\label{AppB}
\subsection{Analytical linearized approach}
In the opposed limit of strong pump amplitude where the cavity fields are dominantly classical, it is convenient to resort to standard linearization techniques allowing to check the convergence of the master equation results. The operators are expanded as $\hat a_j = \alpha_j + \delta \hat a_j$ given ${\alpha _j} = \left\langle {{{\hat a}_j}} \right\rangle$ the classical fields amplitudes and $\delta \hat a_j$ the quantum fluctuations operators. Dropping terms of orders larger than 2 in $\delta \hat a_j$ in Eq.(\ref{H}) we can derive a set of linearized quantum Langevin equations
\begin{eqnarray}
\label{a1t} i{\dot{\hat a}_1} &=& \left[ {-{\Delta _1} + 4{U_1}{{\left| {{\alpha _1}} \right|}^2} - i\frac{{{\kappa _1}}}{2}} \right]{{\hat a}_1}\\
\nonumber &+& 2 U_1\alpha _1^2\hat a_1^\dag + i\sqrt {{\kappa _1}} \hat a_1^{\rm{in}}\\
\label{a2t} i{\dot{\hat a}_2} &=& \left[ {-{\Delta _2} + 4{U_2}{{\left| {{\alpha _2}} \right|}^2} - i\frac{{{\kappa _2}}}{2}} \right]{{\hat a}_2}\\
\nonumber &+& 2 U_2\alpha _2^2\hat a_2^\dag + i\sqrt {{\kappa _2}} {\tilde{ a}}_2^{\rm{in}}
\end{eqnarray}
for the fluctuations where we have dropped the $\delta$ notation. The operators $\hat a_1^{\rm{in}}$ and $\tilde a_2^{\rm{in}}$ account for the input noise in each cavities
Here the cavity 1 output $\hat a_1^{\rm{out}} = \sqrt{\kappa_1}\hat a_1 + \hat a_1^{\rm{in}}$ is driving the cavity 2 input ${\tilde{ a}}_2^{\rm{in}}=\sqrt{\eta}\hat a_1^{\rm{out}} + \sqrt{(1-\eta)}\hat a_2^{\rm{in}}$ when an undelayed coupling is assumed and where $\hat a_2^{\rm{in}}$ is the local noise acting on cavity 2. The parameter $\eta\in[0,1]$ stands for the fraction of cavity 1 field that is transmitted to the cavity 2. Hence we obtain
\begin{eqnarray}
\label{a12t} i{\dot{\hat a}_1} &=& \left[ {-{\Delta _1} + 4{U_1}{{\left| {{\alpha _1}} \right|}^2} - i\frac{{{\kappa _1}}}{2}} \right]{{\hat a}_1}\\
\nonumber &+& 2 U_1\alpha _1^2\hat a_1^\dag + i\sqrt {{\kappa _1}} \hat a_1^{\rm{in}}\\
\label{a22t} i{\dot{\hat a}_2} &=& \left[ {-{\Delta _2} + 4{U_2}{{\left| {{\alpha _2}} \right|}^2} - i\frac{{{\kappa _2}}}{2}} \right]{{\hat a}_2}\\
\nonumber &+& 2 U_2\alpha _2^2\hat a_2^\dag + i\sqrt {\eta{\kappa _1\kappa _2}}\hat a_1\\
\nonumber &+& i\sqrt {\eta{\kappa _2}} \hat a_1^{\rm{in}} + i\sqrt{(1-\eta)\kappa_2}\hat a_2^{\rm{in}}
\end{eqnarray}
which reveal squeezing terms of strength $U_j\alpha_j^2$ governed by the steady state classical fields that fulfill
\begin{eqnarray}
\label{alpha1}  0 &=& \left[ {-{\Delta _1} + 2{U_1}{{\left| {{\alpha _1}} \right|}^2} - i\frac{{{\kappa _1}}}{2}} \right]{\alpha _1} + F_1\\
\label{alpha2}  \nonumber 0 &=& \left[ {-{\Delta _2} + 2{U_2}{{\left| {{\alpha _2}} \right|}^2} - i\frac{{{\kappa _2}}}{2}} \right]{\alpha _2} + F_2 - i\chi{\alpha _1} \\
\end{eqnarray}
Eqs.(\ref{alpha1t},\ref{alpha2t}) admit cumbersome but analytical solutions that can be plugged into Eq.(\ref{a12t},\ref{a22t}) which can be recast in the form $\dot{\mathbf{u}} = {\hat A}{\mathbf{u}} + \mathbf{n}$ where
\begin{eqnarray}
\label{n} {\mathbf{u}} &=& {( {{{\hat a}_1},\hat a_1^\dag,{{\hat a}_2},\hat a_2^\dag } )^T}\\
{\mathbf{n }} &=& (\sqrt {{\kappa _1}} \hat a_1^{{\text{in}}},\sqrt {{\kappa _1}} \hat a_1^{\dag {\text{in}}},\sqrt {\eta {\kappa _2}} \hat a_1^{{\text{in}}} + \sqrt {\left( {1 - \eta } \right){\kappa _2}} \hat \nonumber a_2^{{\text{in}}},\\
\label{eta}
&& \sqrt {\eta {\kappa _2}} \hat a_1^{\dag {\text{in}}} + \sqrt {\left( {1 - \eta } \right){\kappa _2}} \hat a_2^{\dag {\text{in}}}{)^T}
\end{eqnarray}
and
\begin{equation}\label{A}
\hat A =  - i\left( {\begin{array}{*{20}{c}}
  {{{\tilde \Delta }_1}}&{2{U_1}\alpha _1^2}&0&0 \\
  { - 2{U_1}\alpha _1^{2*}}&{ - \tilde \Delta _1^*}&0&0 \\
  {-i\chi }&0&{{{\tilde \Delta }_2}}&{2{U_2}\alpha _2^2} \\
  0&{i\chi }&{ - 2{U_2}\alpha _2^{2*}}&{ - \tilde \Delta _2^*}
\end{array}} \right)
\end{equation}
where we used the definition $\tilde \Delta_j={-{\Delta_j} + 4{U_j}{{\left| {{\alpha_j}} \right|}^2} - i\kappa_j/2}$. We can finally write the following Lyapunov equation
\begin{equation}\label{Lyapunov}
  \hat A \hat V + \hat V{\hat A^T} =  - \hat D
\end{equation}
for the steady state correlation matrix
\begin{equation}\label{V}
  \hat V = \left( {\begin{array}{*{20}{c}}
  {\langle {{{\hat a}_1}{{\hat a}_1}} \rangle }&{\langle {{{\hat a}_1}\hat a_1^\dag } \rangle }&{\langle {{{\hat a}_1}{{\hat a}_2}} \rangle }&{\langle {{{\hat a}_1}\hat a_2^\dag } \rangle } \\
  {\langle {\hat a_1^\dag {{\hat a}_1}} \rangle }&{\langle {\hat a_1^\dag \hat a_1^\dag } \rangle }&{\langle {\hat a_1^\dag {{\hat a}_2}} \rangle }&{\langle {\hat a_1^\dag \hat a_2^\dag } \rangle } \\
  {\langle {{{\hat a}_2}{{\hat a}_1}} \rangle }&{\langle {{{\hat a}_2}\hat a_1^\dag } \rangle }&{\langle {{{\hat a}_2}{{\hat a}_2}} \rangle }&{\langle {{{\hat a}_2}\hat a_2^\dag } \rangle } \\
  {\langle {\hat a_2^\dag {{\hat a}_1}} \rangle }&{\langle {\hat a_2^\dag \hat a_1^\dag } \rangle }&{\langle {\hat a_2^\dag {{\hat a}_2}} \rangle }&{\langle {\hat a_2^\dag \hat a_2^\dag } \rangle }
\end{array}} \right)
\end{equation}
Given that in the absence of thermal photons the only non-zero noise correlations are $\langle {\hat a_j^{{\text{in}}}\hat a_j^{\dag {\text{in}}}} \rangle$ the corresponding matrix reads
\begin{equation}\label{D}
\hat D = \frac{1}{2}\left( {\begin{array}{*{20}{c}}
  0&{{\kappa _1}}&0&{\chi} \\
  0&0&0&0 \\
  0&{\chi}&0&{\left( {1 - \eta } \right){\kappa _2}} \\
  0&0&0&0
\end{array}} \right)
\end{equation}
Eq.(\ref{Lyapunov}) can be vectorized in the form
\begin{equation}\label{LyapunovVec}
  \left( {\mathbb{\hat I} \otimes \hat A + \hat A \otimes \mathbb{\hat I}} \right){\rm{vec}}( \hat V ) =  -{\rm{vec}}( \hat D )
\end{equation}
which is nothing but a set of linear equation solvable analytically. Then, from the knowledge of $\hat V$, we can compute the second order coherence functions as
\begin{equation}\label{g2lin}
  g_j^{(2)}\left( 0 \right) \simeq \frac{{N_j^2 + 4 n_j N_j + 2n_j + 2\operatorname{Re} \left[ {\alpha _j^{*2}\langle {\hat a_j^2} \rangle } \right] + {{| {\langle {\hat a_j^2} \rangle }|}^2}}}{{{{\left[ {n_j  + N_j} \right]}^2}}}
\end{equation}
with the definitions $N_j=|\alpha_j|^2$ and $n_j=\langle\hat a_j^\dag\hat a_j\rangle$ for the classical and fluctuation populations respectively. Finally we note that while such linearized approach only allows for gaussian states, it remains suitable for the description of squeezed states.

\subsection{Semiclassical Approach}
\begin{figure}[ht]
\renewcommand{\figurename}{Fig.A}
\includegraphics[width=0.5\textwidth,clip]{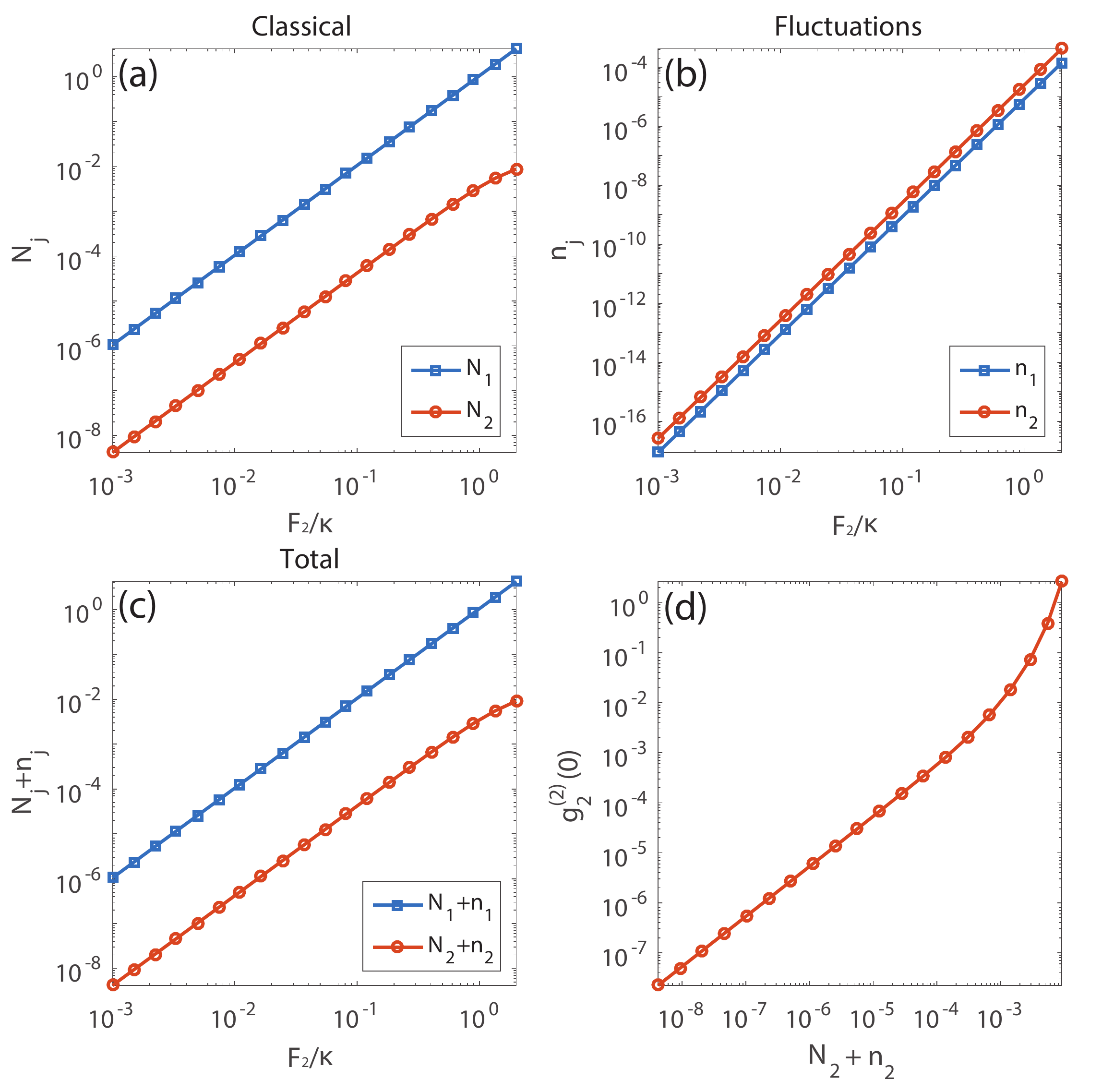}\\
\caption{(Color online) Semiclassical steady state solutions for the parameters of Fig.2 of the main text under the optimal condition. (a) Classical, (b) fluctuations and (c) total cavities occupations versus the pump amplitude $F_2$. (d) Cavity 2 second order coherence function versus its total occupation $N_2+n_2$.}
\label{FigS1}
\end{figure}

While the above linearized approach allows to efficiently compute the steady state solutions and is especially suitable for optimization, it cannot directly address the system dynamics, disregards non-Gaussian states and doesn't give access to the system density matrix. If needed one can therefore deploy a semiclassical approach to overcome these limitations where the classical field dynamics is governed by
\begin{eqnarray}
\label{alpha1t}  i {\dot \alpha}_1 &=& \tilde \Delta_1{\alpha _1} + F_1(t)\\
\label{alpha2t}  i {\dot \alpha}_2 &=& \tilde \Delta_2{\alpha _2} + F_2(t) - i\chi{\alpha _1}
\end{eqnarray}
and the fluctuations are evolving according to the master equation
\begin{eqnarray}\label{rhotf}
  i\frac{{\partial \hat \rho_f }}{{\partial t}} =  \left[ {\hat {\cal{H}}_f,\hat \rho_f } \right] &-& {\frac{{{i}}}{2}\sum\limits_{j = 1,2} \kappa _j\hat {\cal{D}}\left[ {{{\delta\hat a}_j}} \right]\hat \rho_f}\\
\nonumber    &+& i\chi \hat {\cal{D}}\left[ {{{\delta\hat a}_1},{{\delta\hat a}_2}} \right]\hat \rho_f
\end{eqnarray}
where the associated Hamiltonian reads
\begin{eqnarray}
{{\hat {\cal H}}_f} &=& \sum\limits_{j = 1,2} {\left[ {\tilde \Delta_j\hat a_j^\dag {{\hat a}_j} + {U_j}\left( {\alpha _j^{2*}\hat a_j^2 + \alpha _j^2\hat a_j^{\dag 2}} \right)} \right]}\\
\nonumber  &+& \sum\limits_{j = 1,2} {{U_j}\left[ {\hat a_j^\dag \hat a_j^\dag {{\hat a}_j}{{\hat a}_j} + 2\alpha _j^*\hat a_j^\dag {{\hat a}_j}{{\hat a}_j} + 2{\alpha _j}\hat a_j^\dag \hat a_j^\dag {{\hat a}_j}} \right]}
\end{eqnarray}
when dropping the $\delta$ notations. Note that we keep here nonlinear terms of all orders for an exact description of the system. The advantage of this method with respect to the direct master equation treatment stems from the fact that the fluctuation occupation is very small in the displaced reference frame set by the classical field which allows to work with a much smaller truncated Hilbert space and this at an arbitrary pump power. It turns into a great advantage for the global optimization routine that converges much faster and small memory cost in the semiclassical case. If needed, the full system density matrix can be reconstructed applying a multimode displacement operator according to
\begin{eqnarray}\label{Displacement}
    \hat {\cal{D}}(\alpha_1,\alpha_2) &=& {e^{{\alpha _1}\hat a_1^\dag  - \alpha _1^*{{\hat a}_1}}}{e^{{\alpha _2}\hat a_2^\dag  - \alpha _2^*{{\hat a}_2}}} \\
    \hat \rho &=& \hat {\cal{D}}(\alpha_1,\alpha_2)\hat \rho_f\hat {\cal{D}}^\dag(\alpha_1,\alpha_2)
\end{eqnarray}
where $\hat \rho_f$ as been preliminary inflated with zeros to a suitable size imposed by the classical amplitudes. Any correlation can be accessed by reconstructing the global operators $\hat a_j(t) = \alpha_j(t) \mathbb{I} + \delta \hat a_j$ evaluated on $\hat \rho_f(t)$. We show in Fig.A\ref{FigS1} results corresponding to Fig.2(a) of the main text using the above semiclassical approach (see captions).

\section{Appendix C: Optimal pulse delay}\label{AppC}
\begin{figure}[ht]
\renewcommand{\figurename}{Fig.A}
\includegraphics[width=0.5\textwidth,clip]{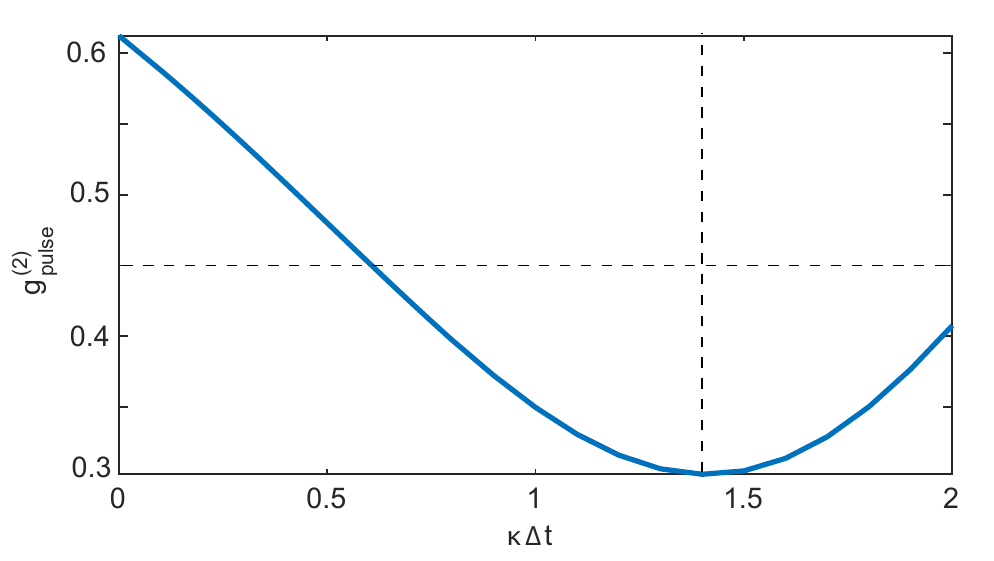}\\
\caption{(Color online) Integrated cavity 2 emission statistics $g^{(2)}_{\rm{pulse}}$ over one pulse at variable delays $\Delta t$ between the 2 cavity pulses.}
\label{FigS2}
\end{figure}

As discussed in the main text, the $g^{(2)}_{\rm{pulse}}$ can be minimized by imposing the proper delay $\Delta t=t_{02}-t_{01}$ between the cavity pulses. We have therefore computed in Fig.A\ref{FigS2} the $g^{(2)}_{\rm{pulse}}$ dependance versus $\Delta t$ for the parameters of Fig.5 which demonstrates the occurrence of an optimal value.

\section{Appendix D: Unidirectional Coupling}\label{AppD}
\begin{figure}[ht]
\renewcommand{\figurename}{Fig.A}
\includegraphics[width=0.5\textwidth,clip]{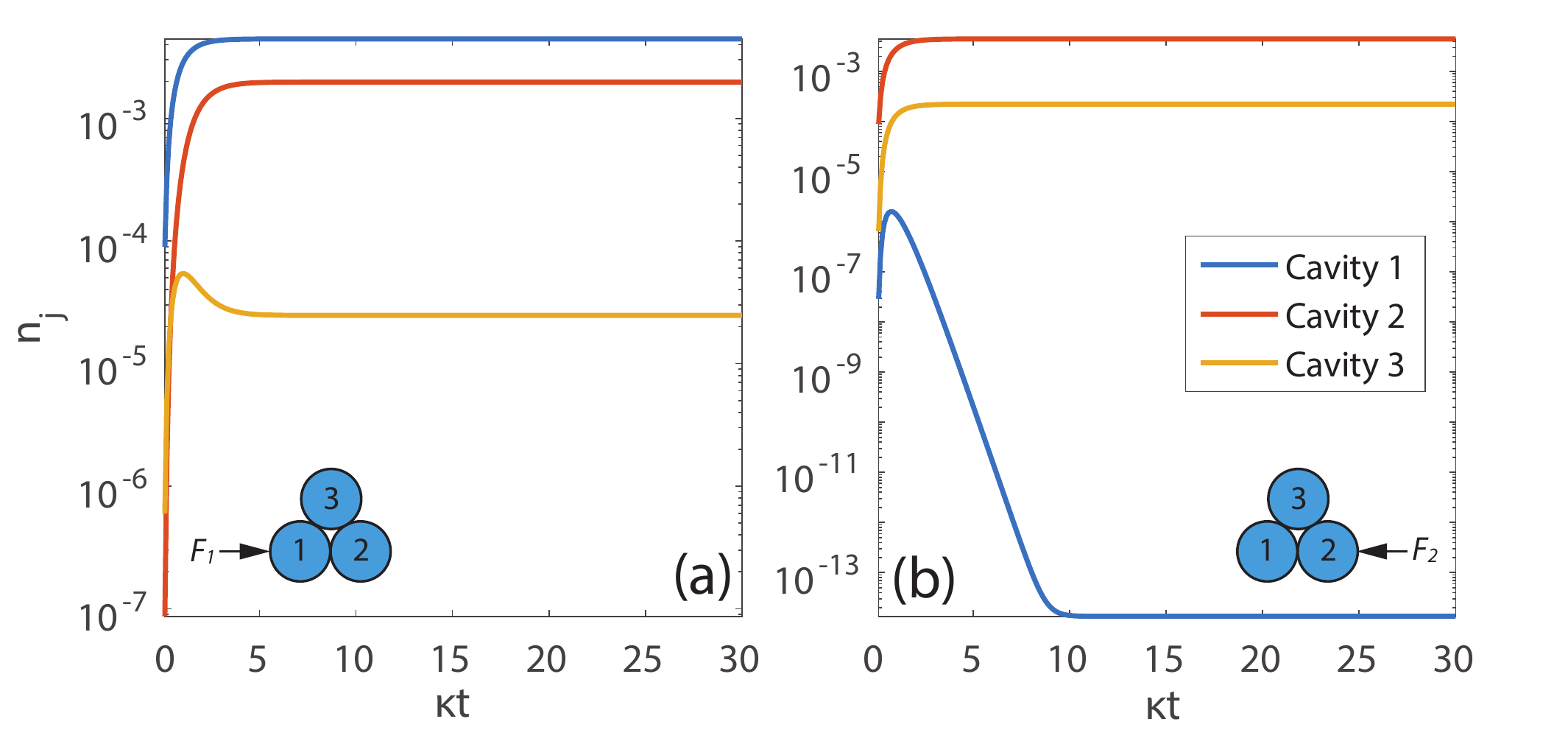}\\
\caption{(Color online) Log scale cavity occupations in the case where we drive (a) only the cavity 1: $F_1=0.1\kappa$ and $F_2=0$ (b) only the cavity 2: $F_1=0$ and $F_2=0.1\kappa$.}
\label{FigS3}
\end{figure}

In the main text we proposed the three cavity configuration as a candidate system for the required dissipative coupling. In that case, the master equation takes the standard form
\begin{equation}\label{rhot3}
  i\frac{{\partial \hat \rho }}{{\partial t}} =  \left[ {\hat {\cal{H}},\hat \rho } \right] - {\frac{{{i}}}{2}\sum\limits_{j = 1}^3 \kappa _j\hat {\cal{D}}\left[ {{{\hat a}_j}} \right]\hat \rho}
\end{equation}
where ${\hat {\cal{H}}}$ is defined in \eqref{H3}. We set the conditions $J_{12} = i{\chi }/{2}$, $J_{23} = J_{31}=\sqrt{-iJ_{12}\kappa_3/2}$, $F_3=0$, $U_1=U_2=U_3=10^{-3}\kappa$, $\kappa_1=\kappa_2=\kappa$ and $\kappa_3=10\kappa$. To demonstrate the unidirectional transmission we solve the system dynamics through Eq.\eqref{rhot3}, in the cases where only one cavity is driven as shown in Fig.A\ref{FigS3} and setting the nonzero pump amplitude to $0.1\kappa$. As one can see, while driving the cavity 1 solely results in a nonzero field in the cavity 2, driving only the latter results in a vanishing cavity 1 occupation as expected. It demonstrates the efficient unidirectional transmission from cavity 1 to cavity 2 in the parameter range we consider.

\section{Appendix E: Indistinguishability}\label{AppE}
\begin{figure}[ht]
\renewcommand{\figurename}{Fig.A}
\includegraphics[width=0.5\textwidth,clip]{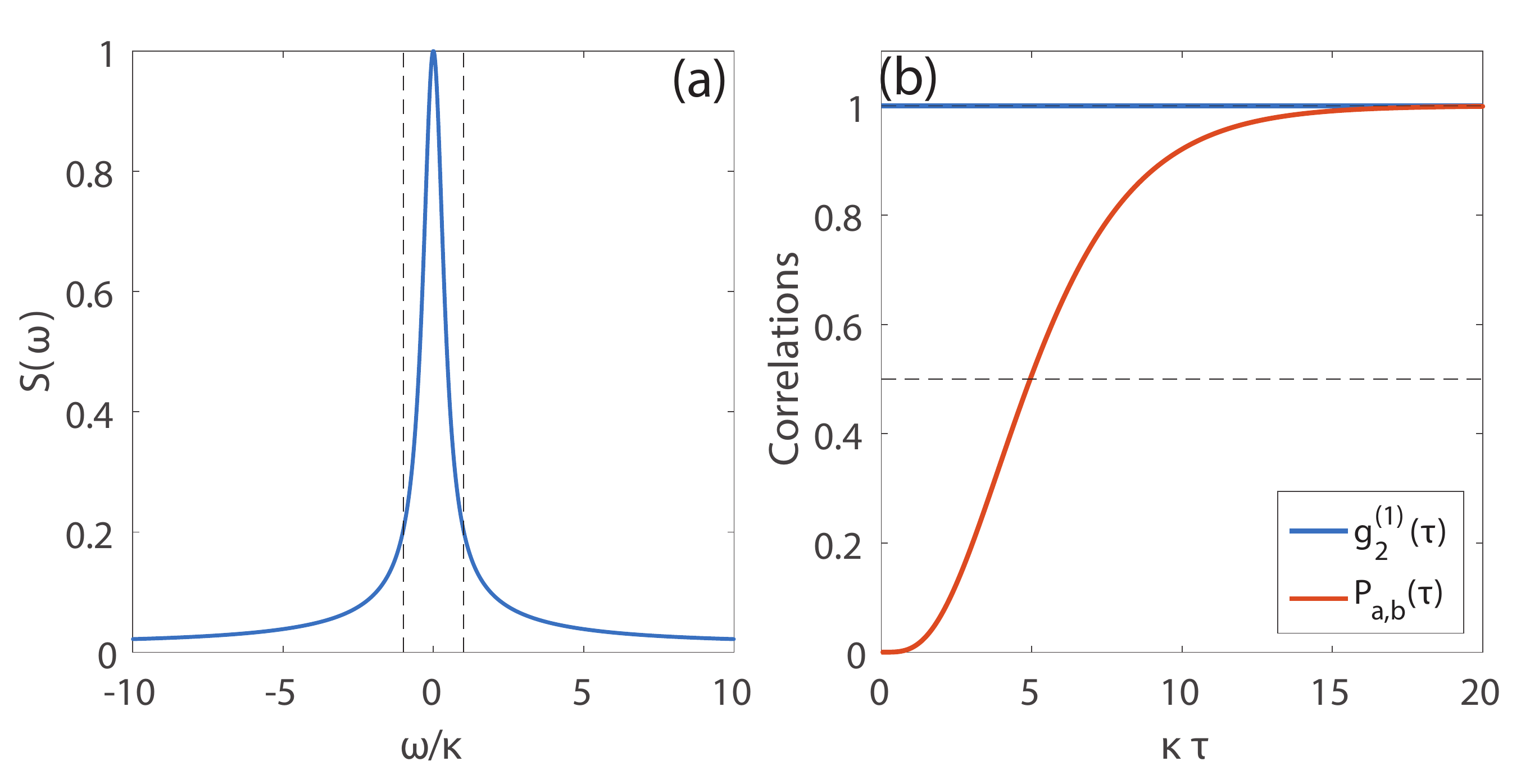}\\
\caption{(Color online) (a) Target cavity spectrum $S(\omega)$. (b) Normalized first order coherence function $g^{(1)}_2(\tau)$ and normalized Hong-Ou-Mandel coincidences out of a beamsplitter in the steady state regime. The driving fields are set as $F_2=0.1\kappa$ and $F_1=F_1|_{\rm{opt}}$.}
\label{FigS4}
\end{figure}

As discussed above, a strong asset of the dissipative coupling is the absence of normal mode splitting that characterizes the UPB. The spectrum $S(\omega)=\int \langle\hat a_2^\dag(0)\hat a_2(\tau) \rangle d\tau$ of the target cavity is shown in Fig.\ref{FigS4}(a) and is nothing but a Lorenzian spectrum of width $\kappa_2$. This strongly favors the indinstinguishability of the single photon emission which can be quantified by simulating a Hong-Ou-Mandel experiment. Let us imagine that the output of the target cavities of two identical cascaded systems $a$ and $b$ are mixed in a $50/50$ beamsplitter. One can show \cite{Woolley2013} that the coincidence probability delayed by $\tau$ between the beamsplitter outputs reads
\begin{equation}\label{P12}
  P_{a,b}(\tau) = \frac{1}{2}\left( G^{(2)}_2(\tau) - |G^{(1)}_2(\tau)|^2 + G^{(1)}_2(0)^2 \right)
\end{equation}
where $G^{(1)}_2(\tau)=\langle \hat a_2^\dag(0)\hat a_2^\dag(\tau)\rangle$ and $G^{(2)}_2(\tau)=\langle \hat a_2^\dag(0)\hat a_2^\dag(\tau)\hat a_2(\tau)\hat a_2(0)\rangle$. Interestingly in the steady state regime, $G^{(1)}_2(\tau)$ is constant as one can see in Fig.\ref{FigS4}(b) (blue line). Therefore $P_{a,b}(\tau)$, shown by the red line of Fig.\ref{FigS4}(b) in its normalized form, is essentially defined by the two-photon correlation $G^{(2)}_2(\tau)$ and in particular $P_{a,b}(0)=G^{(2)}_2(0)$. The indistinguishability degree is quantified by the visibility
\begin{equation}\label{V}
  {{\cal{V}} = \frac{{{P_{a,b}}\left( { + \infty } \right) - {P_{a,b}}\left( 0 \right)}}{{{P_{a,b}}\left( { + \infty } \right) + {P_{a,b}}\left( 0 \right)}}} = 1 - g^{(2)}_2(0)
\end{equation}
which amounts to ${\cal{V}}=99.95\%$ in our case. Within the assumptions of our model, this value is simply determined by the purity ${\cal P}={\rm Tr}(\rho_2^2)$ of the target cavity state and would obviously reach a value of $100\%$ (perfect indistinguishability) for a pure state \eqref{psi}. In the pulsed regime, the reasoning would still hold and to plot $P_{a,b}(\tau)$ in that context, one would have to sum realizations at variable delay between pairs of pulses sent on the beamsplitter \cite{Woolley2013}.

\bibliography{Bibliography}

\end{document}